# Shot Range and High Order Correlations in Proteins

Shiyang Long,[†] and Pu Tian[*,†,‡]

*College of Life Science, and MOE Key Laboratory of Molecular Enzymology and Engineering*

*Jilin University*

*2699 Qianjin Street, Changchun 130012*

E-mail: tianpu@jlu.edu.cn

## Abstract

The main chain dihedral angles play an important role to decide the protein conformation. The native states of a protein can be regard as an ensemble of a lot of similar conformations, in different conformations the main chain dihedral angles vary in a certain range. Each dihedral angle value can be described as a distribution, but only using the distribution can't describe the real conformation space. The reason is that the dihedral angle has correlation with others, especially the neighbor dihedral angles in primary sequence. In our study we analysis extensive molecular dynamics (MD) simulation trajectories of eleven proteins with different sizes and folds, we found that in stable second structure the correlations only exist between the dihedrals near to each other in primary sequence, long range correlations are rare. But in unstable structures (loop) long range correlations exist. Further we observed some characteristics of the short range correlations in different second structures (α-helix, β-sheet) and we found that we can approximate good high order dihedral angle distribution good only use three order distribution in stable second structure which illustrates that high order correlations (over three order) is small in stable second structure.

## Introduction

The empirical distributions of the backbone dihedral angles ϕ and ψ of amino acids in proteins have been studied for over 40 years. And some probability densities are publicly which can be used in structure validation and structure prediction[1]. But it's not a good way to study the correlation of neighbor residues with these distributions, because the data points are limited when we group the data sets as residue pairs and in different protein the residue pairs are in different environment so the correlations are mixed. Molecular dynamics (MD) simulation trajectories can over these problems, we can get enough data sets of each residue pairs and the meaning of the correlation is clear.

Conformational dynamics of folded proteins plays an essential role in biologically important functional processes, such as enzymatic activity, molecular recognition, and allosteric regulation[2, 3]. Correlation play an important role to understand the dynamics. Calculating correlations from MD trajectories has a long history, earlier reaches aim to study the collective motions and local conformational change in proteins with cross correlation of protein atoms[4-6]. The correlations of protein backbone dihedrals are also studied[7-9]. Li *et. al.*[7] analyzed a 700-ns molecular dynamics (MD) simulation trajectory of ubiquitin and calbindin $D_{9k}$, and concluded that long-range pair correlations are rather rare and all of dihedral angle pairs with large correlation are at short range. Though dihedral short range correlations are observed by us, it is not researched carefully. Here we will research the short range backbone dihedral correlations in different second structures using eleven MD trajectories.

## Result and Method

Here we use eleven MD trajectories, most of them are 1-10 us, cdk2 and lysozyme are longer. Cdk2 is generated using amber force field and other proteins are generated using charmm force field. First we calculate the correlation (mutual information) matrix of each protein. Each backbone dihedral angle is divided into 360 sections equally and the mutual information is calculated for each two dihedral angles. Two matrixes are list in Fig. 1. We can see that on the diagonal line and its neighboring region the correlations are large; they are dihedral angles near to each other in primary sequence. In other regions correlations are rare and most of this kind of correlations are from loop regions. To see the short range correlation clearly we plot the neighbor correlation of 1bta (pdb id) in Fig. 2. There are two kinds of neighbor dihedral angle pairs in protein backbone. One kind is $\phi_i$ and $\psi_i$ pairs we defined fp, another is $\phi_{i+1}$ and $\psi_i$ pairs we define pf. In Fig 2 we divide the dihedral pairs in different second structures (α-helix, β-sheet, others) and DSSP is used here to identify second structure. From the fig we can see different second structure has their own characteristics, in α-helix the correlations are close and pf correlations are a little higher than fp, in β-sheet fp correlations are very low and pf correlations are high, in other structures (such as loops) correlations are not stable some very large correlation exist. To test if all the proteins have these characteristics we plot the distributions of neighbor correlations of different second structure with the data of eleven proteins (Fig. 3). The distributions agree with the results we get in 1bta. So large correlations exist in protein no mater $\phi_i$ and $\psi_i$ dihedral pairs or $\phi_{i+1}$ and $\psi_i$ dihedral pairs, but most of the probability densities only care the $\phi_i$ and $\psi_i$ correlations, to improve the accuracy considering $\phi_{i+1}$ and $\psi_i$ correlations is necessary.

The results from charmm and amber trajectories are similar, but we found that α-helix pf dihedral correlations are higher in cdk2 which is simulation by amber force field. So we simulate lysozyme trajectory with amber force field and compare the results with its charmm trajectory (Fig. 4). We can see that the green points have large values in amber force field result, this further illustrate that in amber force field the α-helix pf dihedral pairs have larger correlation than in charmm. But the reason is not known. Except that difference the correlation characteristics are similar in two force field trajectories. To some extent, it proved that these characteristics may exist in real protein and are not caused by the differences of force field.

We generate a 2 ns lysozyme trajectory and a 200 us trajectory using charmm force field, choosing equal data points from the two trajectories we calculate the short time scale correlation and long time scale correlation. We minus this two kinds of neighbor correlations and plot the results in Fig. 5. We find that most neighbor correlation in α-helix change small in different time scale, so maybe correlations at these stable regions are caused by short time scale dynamics. A small number of correlation differences in α-helix are big, they exist on the edges regions of their structures and they may not be stable second structures.

In previous work high order correlations in proteins are not well studied. Here we will test the importance of high order correlations in another sight. We approximate the high order distributions (six order) with one, two, three order distributions. We divide each dihedral angle into six sections and we make sure each section has same data points. We define the data have the same conformation if they fall in the same section. So we have 46656 conformations if we use six dihedral angles. Each conformation's probability can be approximate by equations (Eq. (1) (2) (3)), if no correlations exist the distribution of the conformations is equal equivalent distribution.

$$p(x_1x_2\ldots x_n) \approx p(x_1)*p(x_2)*\ldots*p(x_n) \quad (1)$$

$$p(x_1x_2\ldots x_n) \approx p(x_1)*p(x_2|x_1)*\ldots*p(x_n|x_{n-1}) \quad (2)$$

$$p(x_1x_2\ldots x_n) \approx p(x_1)*p(x_2|x_1)*p(x_3|x_1x_2)*\ldots*p(x_n|x_{n-2}x_{n-1}) \quad (3)$$

$p(x_1x_2\ldots x_n)$ is the real probability, $p(x_i)$ is the probability of one angle conformation, p(x|y) is the conditional probability. With the influence of correlations equations (Eq. (1) (2) (3)) can't describe the real distribution, but with the conditional probability add the result will be better. We use correlation coefficient of real distribution and approximate distribution to describe the similarity of two distributions, using Eq. (1) the distribution is equivalent distribution so the correlation coefficient is nearly zero. The results list in Fig. 6, we choose different positions of different proteins with the same second structure and in each position we can get a correlation coefficient and a distribution of the correlation coefficient is generated. We can approximate using equations (Eq. (1) (2) (3)), so we get three distributions of each second structure. The fig shows us that we can approximate good the real distributions using three order correlation in α-helix and β-sheet, but at other structures some positions can't get good results. This show us in stable second structures such as α-helix and β-sheet long range and high order (over three) correlations is not much, and in some unstable regions these correlations may exist and we can't neglect them. We also approximate eight order distributions and the result is similar.

We observed short range correlations, but how the correlations generate is not known. So we plot two dimensional distributions of different second structures. We choose an example of each kinds second structure and plot them in Fig. (7)(8)(9). It's clear that most of the short range correlations come from the dynamic that the neighbor dihedral angles move in opposite directions. But in β-sheet fp regions where the correlations are small the dynamic is different from others.

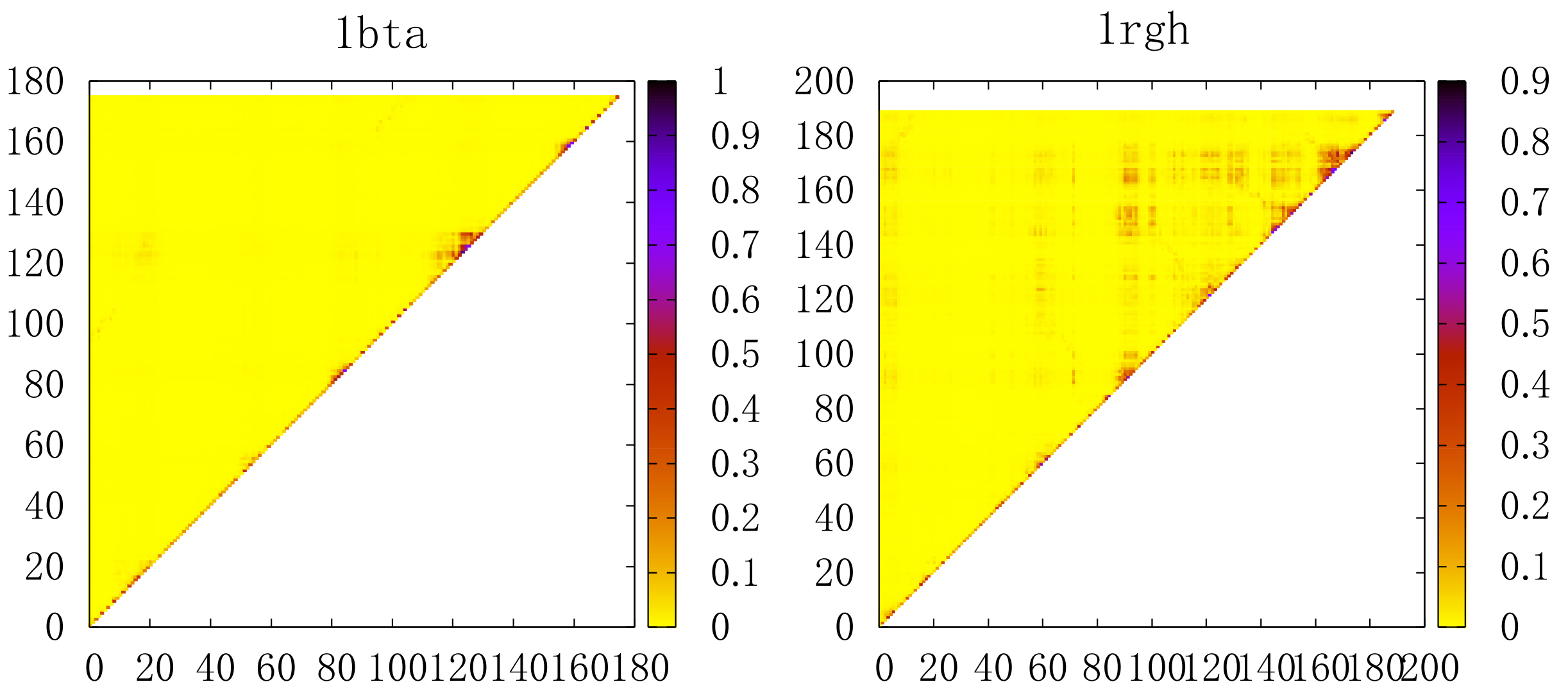

Figure1: 1bta and 1rgh is pdb id of protein crystal structures, each point in the picture present a correlation of a pair of dihedral angles, its x value and y value is the serial number of dihedrals, the color of the points present the correlation value.

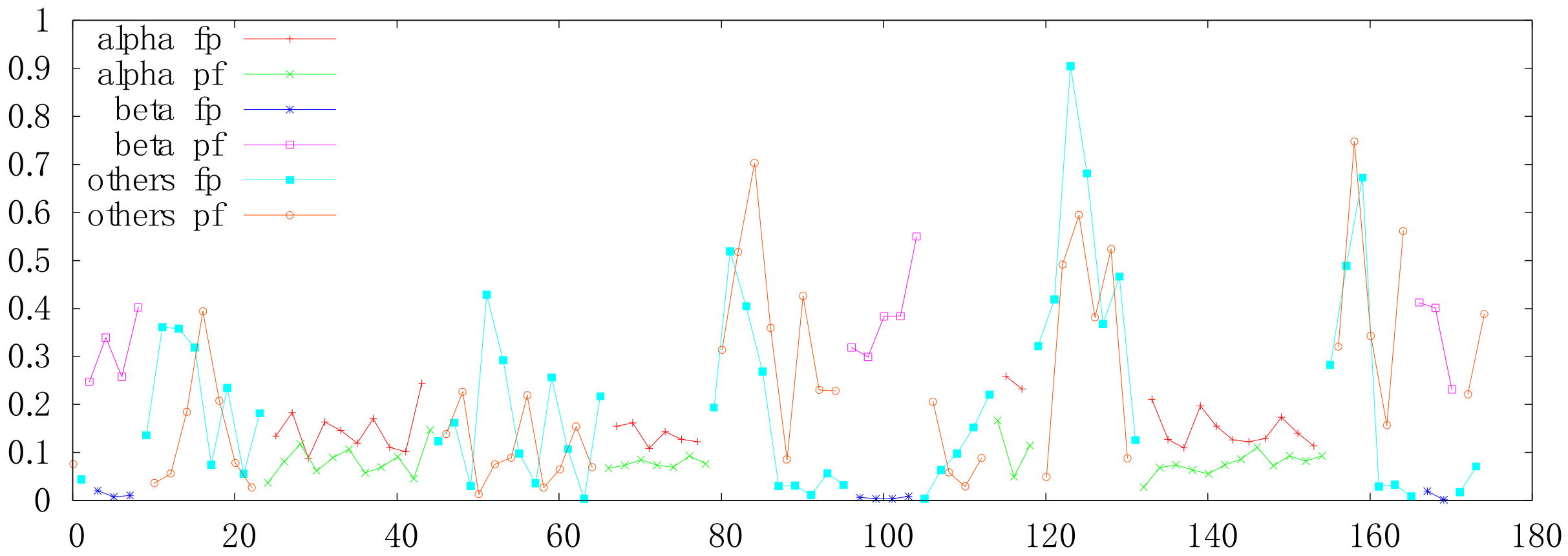

Figure2: Each point in the picture is a correlation of two neighbor dihedral angles, x value is the serial number of first dihedral of the dihedral pair, y value is the correlation value. Alpha is α-helix, beta is β-sheet, fp is $\phi_i$ and $\psi_i$ pairs, pf is $\phi_{i+1}$ and $\psi_i$ pairs.

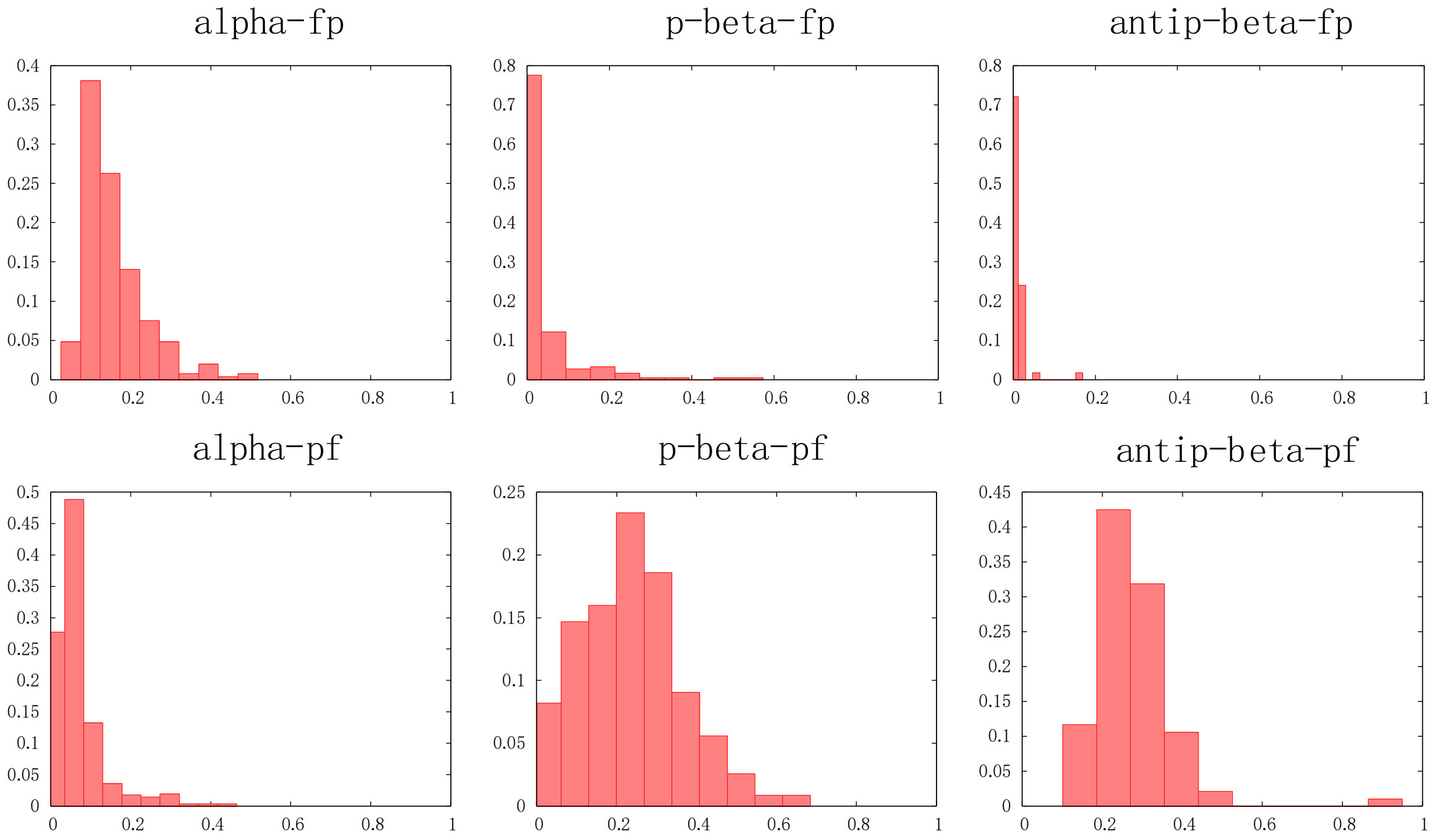

Figure3: Alpha is α-helix, p-beta is parallel β-sheet, antip-beta is antiparallel β-sheet, fp is $\phi_i$ and $\psi_i$ pairs, pf is $\phi_{i+1}$ and $\psi_i$ pairs. X axis is the correlation value; y axis is probability.

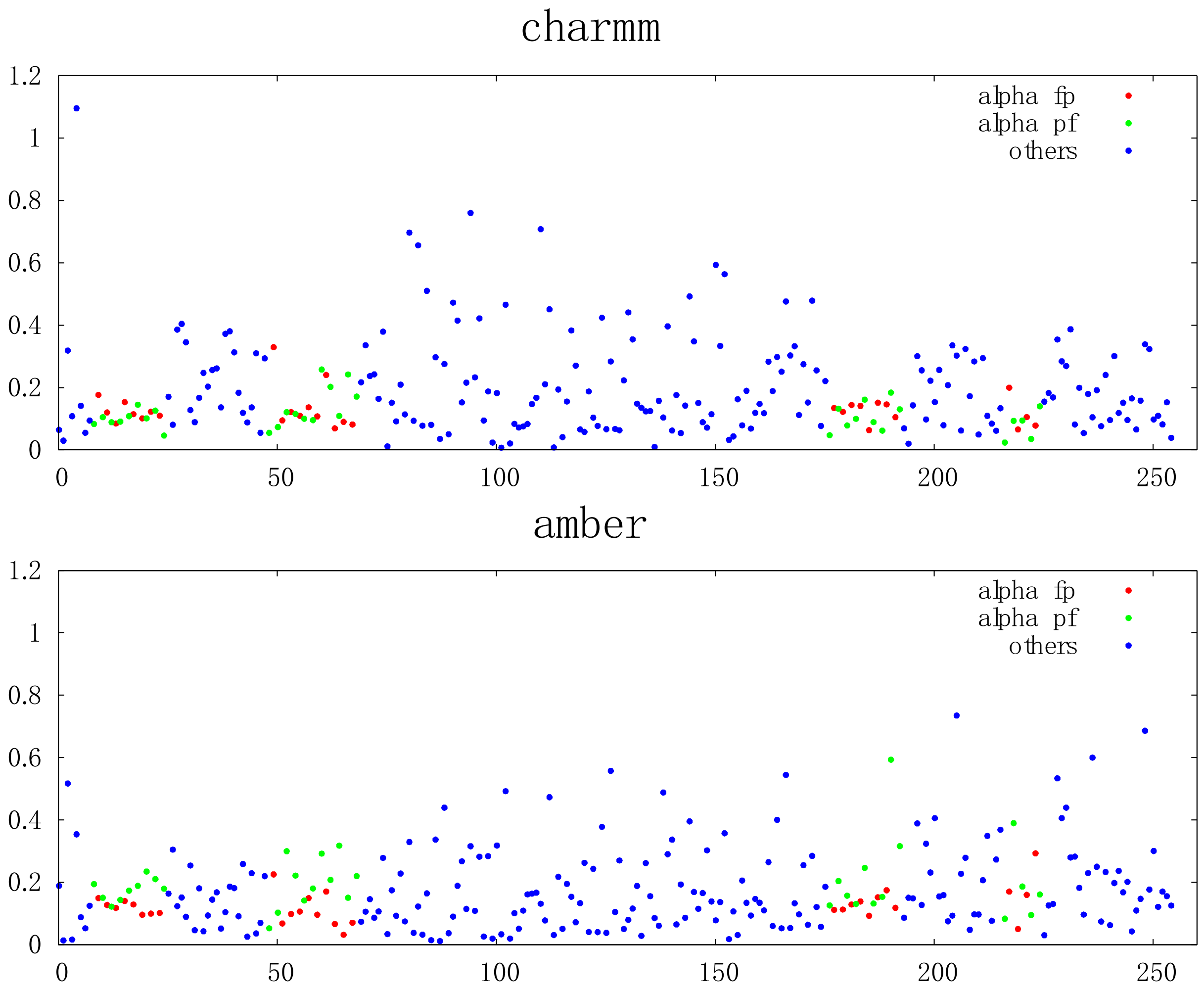

Figure4: Each point in the picture is a correlation of two neighbor dihedral angles, x value is the serial number of first dihedral of the dihedral pair, y value is the correlation value. Alpha is α-helix, others are the other structures, fp is $\phi_i$ and $\psi_i$ pairs, pf is $\phi_{i+1}$ and $\psi_i$ pairs.

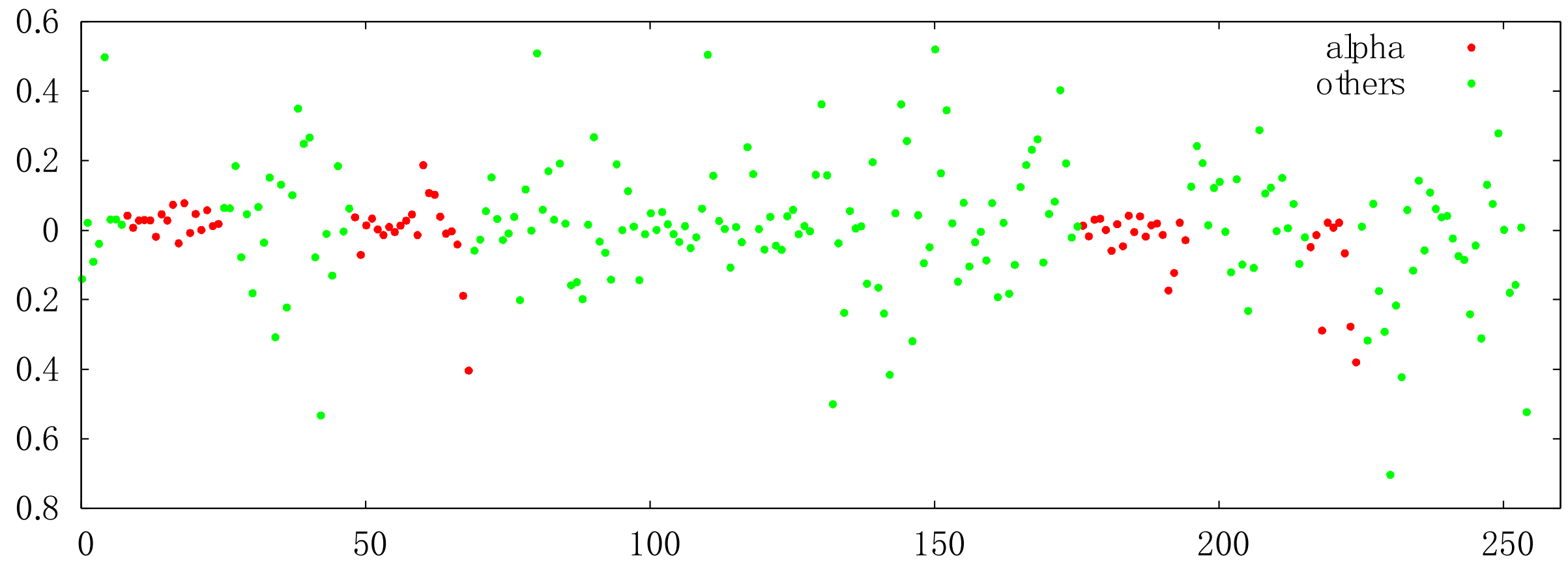

Figure5: X axis is the serial number of first dihedral of the neighbor dihedral pair, y axis is the difference of long time scale correlation and short time scale correlation, alpha is α-helix.

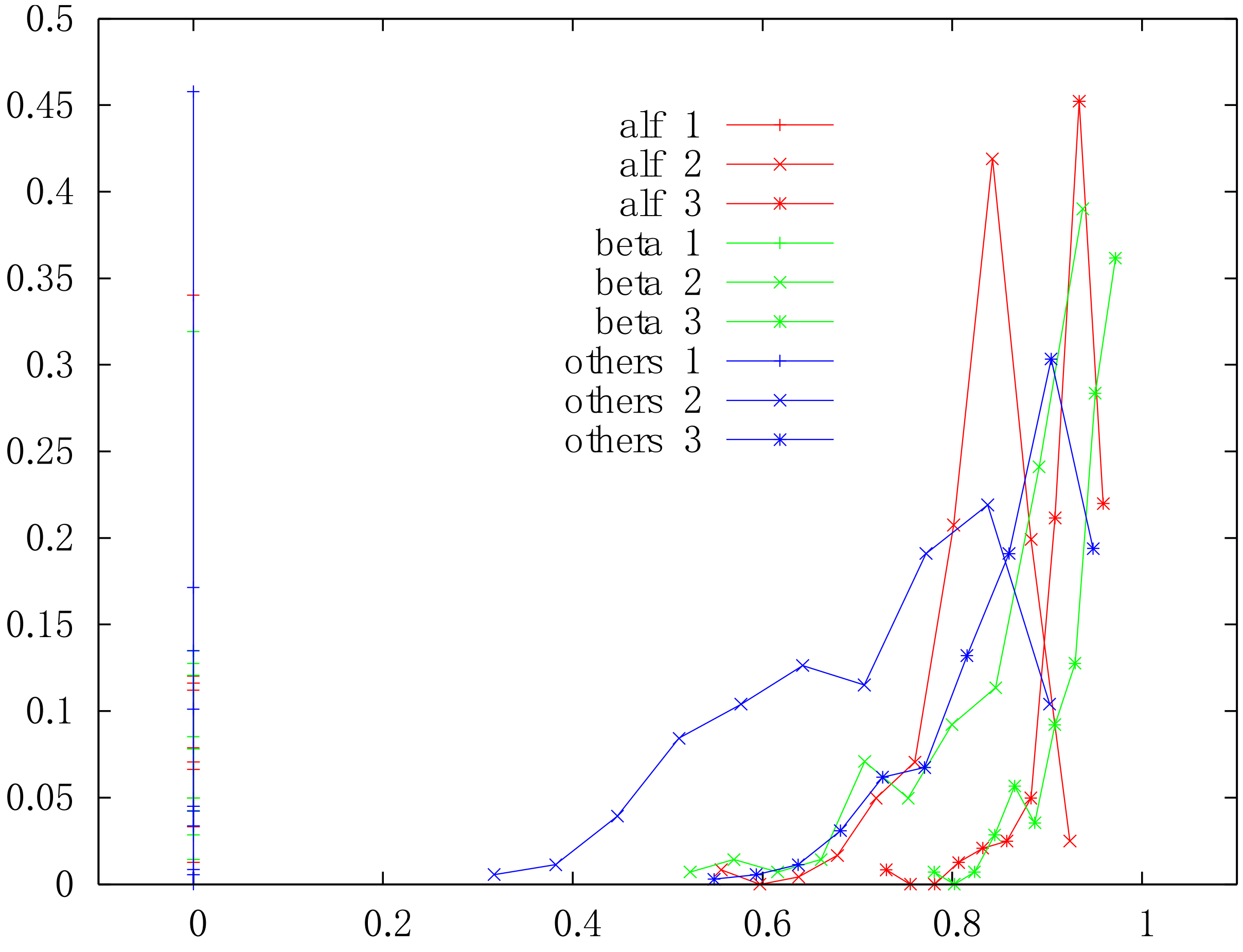

Figure6: X axis is correlation coefficient of real distribution and approximate distribution; y axis is probability of the correlation coefficient. Alf is α-helix, beta is β-sheet, others are other structures, 1 represent approximate with Eq. (1), 2 represent approximate with Eq. (2), 3 represent approximate with Eq. (3).

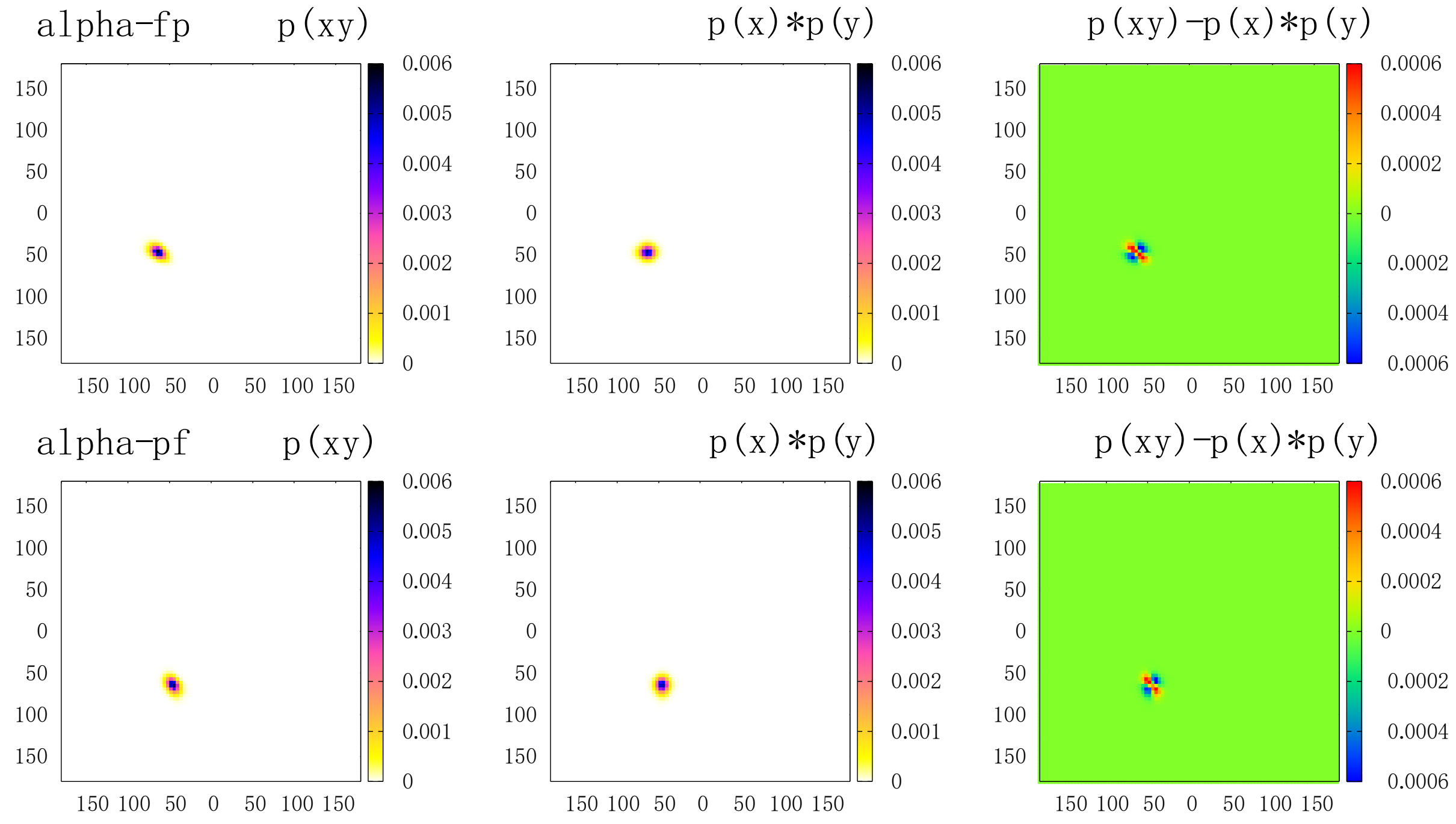

Figure7: Two dimensional distributions of neighbor dihedral angles. A α-helix $\phi_i$ $\psi_i$ pair and a $\phi_{i+1}$ $\psi_i$ pair is choosing here. P(xy) is the real two dimensional distributions, p(x)*p(y) is the distributions calculated with Eq. (1) which canton no correlations. P(xy)-p(x)*p(y) is the differences of the two distributions.

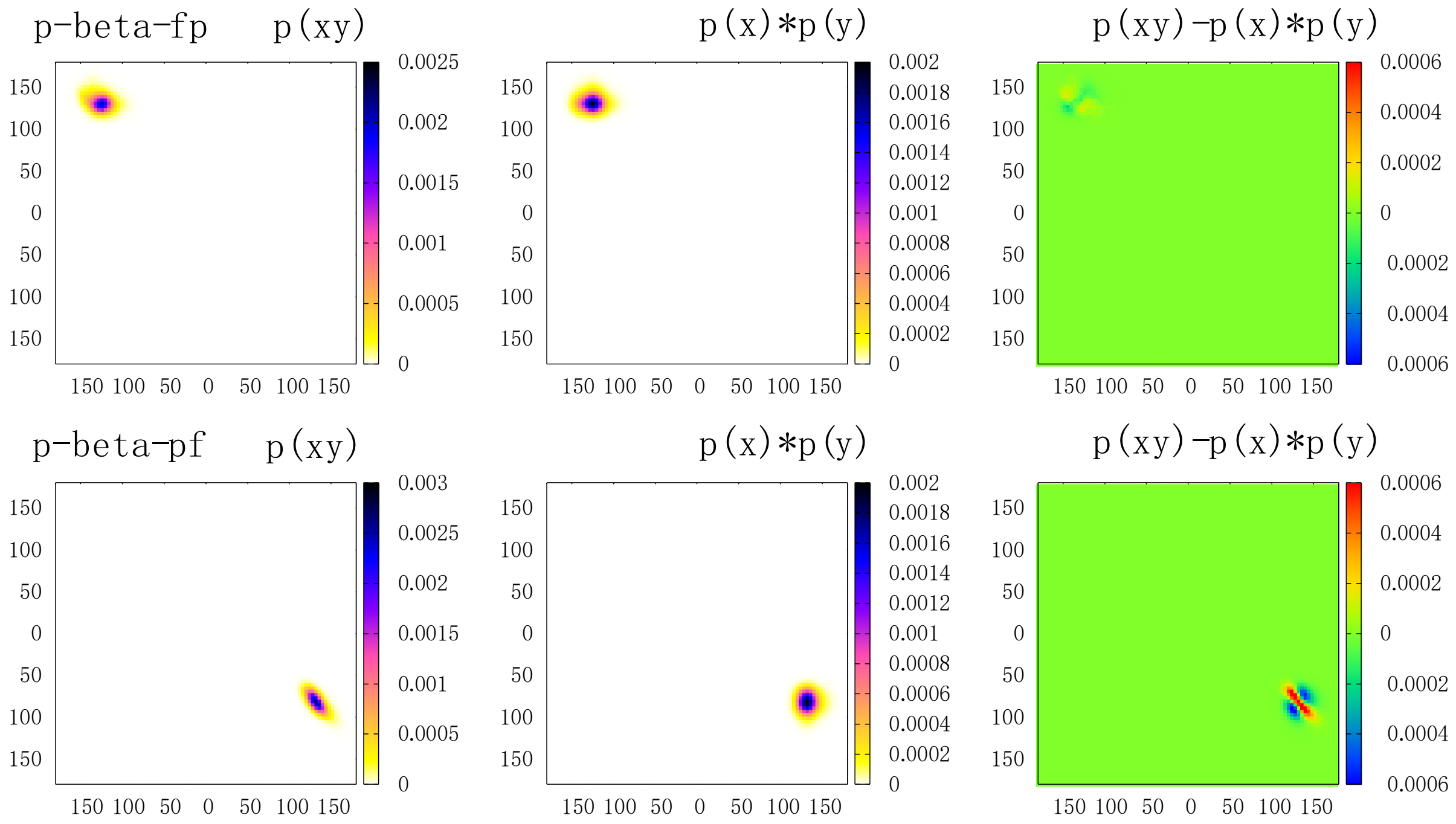

Figure8: Two dimensional distributions of neighbor dihedral angles. A parallel β-sheet $\phi_i$ $\psi_i$ pair and a $\phi_{i+1}$ $\psi_i$ pair is choosing here. P(xy) is the real two dimensional distributions, p(x)*p(y) is the distributions calculated with Eq. (1) which canton no correlations. P(xy)-p(x)*p(y) is the differences of the two distributions.

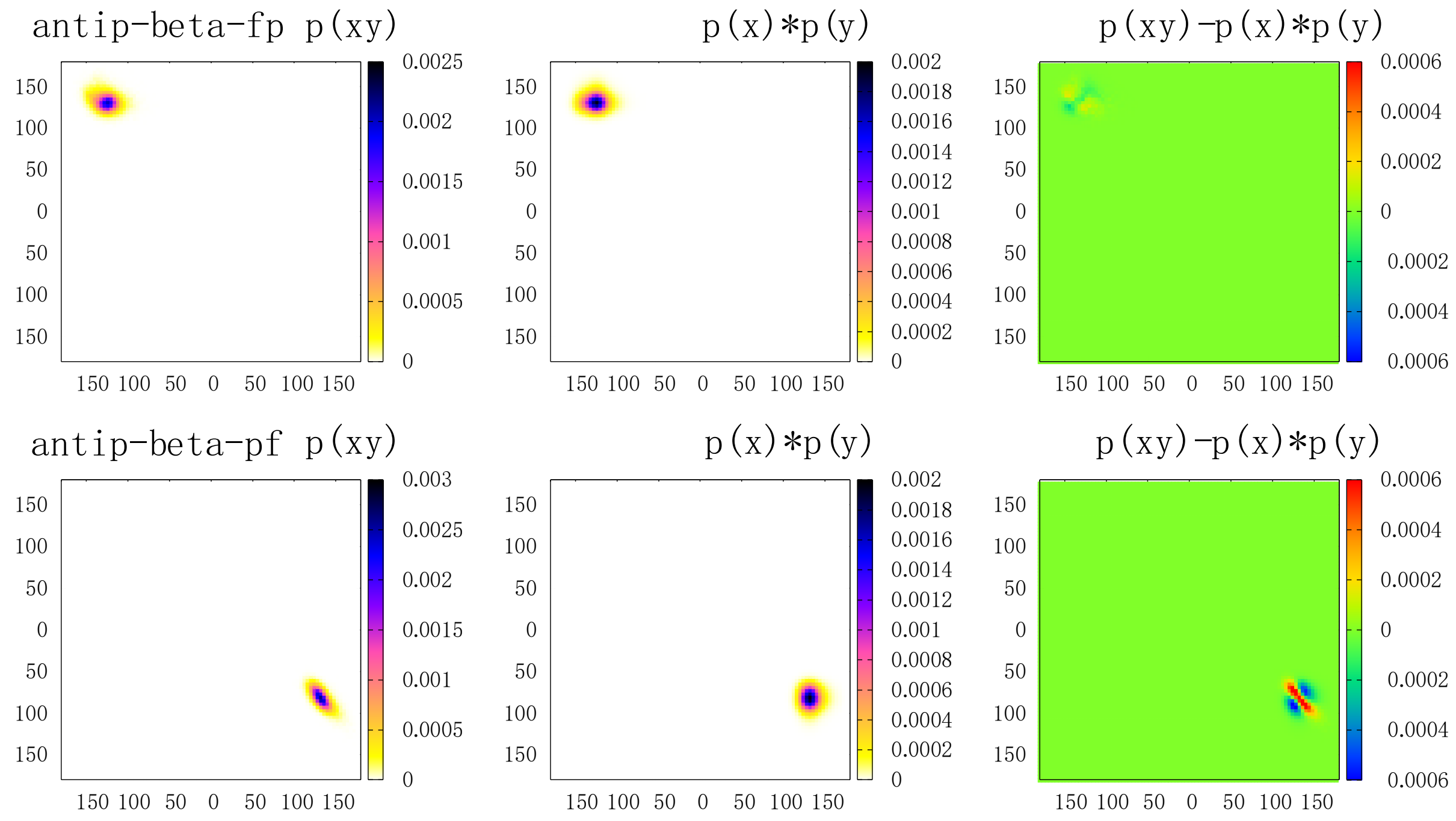

Figure9: Two dimensional distributions of neighbor dihedral angles. An antiparallel β-sheet $\phi_i$ $\psi_i$ pair and a $\phi_{i+1}$ $\psi_i$ pair is choosing here. P(xy) is the real two dimensional distributions, p(x)*p(y) is the distributions calculated with Eq. (1) which canton no correlations. P(xy)-p(x)*p(y) is the differences of the two distributions.